 \documentclass[prd, showpacs]{revtex4}

\usepackage{dcolumn}
\usepackage{amsmath}

\makeatletter
\def\btt#1{\texttt{\@backslashchar#1}}%
\DeclareRobustCommand\bblash{\btt{\@backslashchar}}%
\makeatother

\begin{document}

%\preprint{IITD-DR/CSPA-1}

\title{Exact solutions to the Chandrasekhar Page angular equation }

\author{D. Ranganathan}
 \thanks{Department of Physics, Indian Institute of Technology Delhi, New Delhi 110016 India}

\email{dilip@physics.iitd.ac.in}
\affiliation{
Department of Physics, Indian Institute of Technology Delhi, New Delhi 110016 India}
\date{\today}
\begin{abstract}
Exact solutions are found for the Chandrasekhar Page angular equation which results when the Dirac equation in a Kerr Newman space time is separated into its radial and angular parts. The solutions turn out to be remarkably simple in form while satisfying the asymptotic conditions deduced earlier. The eigenvalues are found to be the square root of the total angular momentum as first found by Dirac for flat space; supplemented by a term which is the product of the mass of the Dirac particle times the specific angular momentum  of the black hole. The additional contribution is what is expected from frame dragging.

\end{abstract}

\pacs{03.65.Ge, 04.20.Jb, 03.65.Pm}

\maketitle

\section{Introduction} \label{Intro:level1}\protect

The ordinary differential equations which result from the separation of the Dirac equation in a Kerr (Kerr-Newman) space time were first derived by Chandrasekhar and Page \cite{chandra:1, page:1, chandra:2}. The Kerr Newman line element in Boyer Lindquist coordinates is given by \cite{chandra:2}
\begin{subequations}
\label{metric}
\begin{eqnarray}
(ds)^2 & = & \frac{\Delta}{\Sigma} (dt - a \: {\sin}^2\theta d\phi)^2 - \Sigma (\frac{{d r}^2}{\Delta} + {d \theta}^2 ) - \frac{{\sin}^2 \theta}{\Sigma} ( a d t - (r^2 + a^2 ) d\phi)^2 ,\\
\Sigma & = & r^2 + a^2, \\
\Delta & = & r^2 - 2 M r + a^2 + Q^2 .
\end{eqnarray}
\end{subequations}

Here $M$ is the black hole mass, $Q$ the black hole's electric charge and $a$ is the hole's angular momentum per unit mass. We will only consider the case $M^2 > Q^2 + a^2$ and work in natural units $G = 1$, $c = 1$ and $\hbar =1$. As this metric is axially symmetric, the Dirac equation can then be separated by writing the four components of the spinor wave function in the form,
\begin{subequations}
\label{psi}
\begin{eqnarray}
{\Psi}_1 & = & \frac{R_{-\frac{1}{2}}(r) \: S_{-\frac{1}{2}}(\theta )}{r - i a \cos\theta}\: e^{i(m \phi - \omega t)} ,\\ 
{\Psi}_2 & = &  R_{\frac{1}{2}}(r) \:   S_{\frac{1}{2}}(\theta )\: e^{i(m \phi - \omega t)},\\ 
{\Psi}_3 & = &   R_{\frac{1}{2}}(r) \: S_{-\frac{1}{2}}(\theta )\:  e^{i(m \phi - \omega t)},\\
{\Psi}_4 & = &  \frac{R_{-\frac{1}{2}}(r) \: S_{\frac{1}{2}}(\theta )}{r + i a \cos\theta}\: e^{i(m \phi - \omega t)}. 
\end{eqnarray}
\end{subequations}

The notation follows Chandrasekhar's book \cite{chandra:2}; slightly modified to use $\omega$ for the frequency (energy) rather than $\sigma$. The $R_{\pm\frac{1}{2}}$ are functions of $r$ alone,  while $S_{\pm\frac{1}{2}}$ are functions of $\theta$. If this form is substituted into the Dirac equation,  we get the following angular equations,
\begin{subequations}
\label{theta-full}
\begin{eqnarray}
\frac{d}{d \theta} S_{+\frac{1}{2}} + \left[ (\frac{1}{2}) \cot\theta + ( a \omega  \sin \theta + m \csc\theta )\right] S_{+ \frac{1}{2}} & = &  ( - \lambda + a m_e \cos\theta) S_{- \frac{1}{2}}, \\
\frac{d}{d \theta} S_{-\frac{1}{2}} + \left[ (  \frac{1}{2}) \cot\theta - ( a \omega  \sin \theta + m \csc\theta ) \right] S_{- \frac{1}{2}}  & = & ( \lambda + a m_e \cos\theta) S_{+ \frac{1}{2}}. 
\end{eqnarray}
\end{subequations}

Again for the radial equations it is convenient to use a slightly different variable from that used in \cite{chandra:2}, we define $ {R^{'}}_{\frac{1}{2}} = \sqrt\Delta R_{\frac{1}{2}}$ to obtain,
\begin{subequations}
\label{rad-full}
\begin{eqnarray}
\frac{d}{d r} {R^{'}}_{+\frac{1}{2}}  + i \frac{\omega (r^2 + a^2) + eQr + m a}{\Delta} {R^{'}}_{+\frac{1}{2}}  & = &  \frac{\lambda - i m_e r}{\sqrt{\Delta}} R_{-\frac{1}{2}} , \\
\frac{d}{d r} R_{-\frac{1}{2}}  -i \frac{\omega (r^2 + a^2) + eQr + m a}{\Delta} R_{-\frac{1}{2}}   & = & \frac{\lambda + i m_e r}{\sqrt{\Delta}} {R^{'}}_{+\frac{1}{2}} .
\end{eqnarray}
\end{subequations}

The separation constant $\lambda$ was shown by Carter and McLenaghan \cite{carter:1} to be the eigenvalue of square root of the total angular momentum operator in the asymptotic limit. This was recently confirmed in an analysis of the Chandrasekhar separation ansatz  by Batic and Schmid \cite{batic:1}.\\

These two first order equations for the angle (\ref{theta-full}), can be combined into a second order equation, which is known as the angular Chandrasekhar Page equation,
\begin{eqnarray}
\label{cs-page-angle}
\frac{1}{\sin \theta} \frac{d}{d \theta}
	\left( \sin \theta \frac{d S}{d \theta}\right)+ \frac{a m_e \sin \theta }{\lambda + a m_e \cos \theta} \frac{d S}{d \theta}  +
\left[\left(\frac{1}{2}- a \omega \cos \theta \right)^2- 
\left(\frac{m - \frac{1}{2} \cos \theta}{\sin \theta}\right)^2 \right.\\  \nonumber
\left.
 -\frac{3}{4} - 2 a \omega m - { a^2 \omega}^2  
+ \frac{ a m_e \left( \frac{1}{2} \cos \theta -\omega a {\sin}^2 \theta - m \right) }{\lambda + a m_e \cos \theta} - a^2 {m_e}^2 {\cos}^2 \theta + {\lambda}^2 \right] S =0.  
\end{eqnarray}

The equations (\ref{theta-full}) and (\ref{cs-page-angle}) have been the object of study since their discovery \cite{fackerell:1, kalnins:1, chakra:1, batic:2}. In the above references, recurrence relations for the series solutions as well as series for the eigenvalues have been found. The numerical nature of these has meant that the solutions are approximate and there have been disagreements as to the actual numerical values \cite{batic:2}. In addition, due to the very large number of parameters; three for the black hole $M, \; Q, \; a$, two for the Dirac particle $m_e,\; e$ and two for the dynamical system as a whole $\omega,\; m$; the possible solution space is very large. In the references cited above numerical results have been found only for a few cases.\\

The full Dirac equations have also been studied by Finster and others \cite{finster:1} and Dafermos \cite{dafermos:1} who claim that no solutions exist for the Dirac equations in this case. For the extremal case $M^2 = Q^2 + a^2$, Schmid \cite{schmid:1} has found exact solutions. The system has also been studied in integral equation form by Guihao Tian \cite{tian:1}. Many of these workers expand the solution in the orthonormal basis formed by the spin weighted spheroidal harmonics \cite{newman:1, goldberg:1, staru:1, berti:1}. This appears to be the natural basis to expand as in the limit $M = 0, \; Q = 0$, the metric eq.(\ref{metric}) reduces to the Minkowski metric in oblate spheroidal coordinates. The Dirac equation is separable in this coordinate system and the angular solutions are spin weighted spheroidal harmonics. If further $ a = 0$ also, these reduce to spin weighted spherical harmonics \cite{ newman:1, goldberg:1}.\\

The emphasis in this work is on the requirement of a finite probability density everywhere for the wave function. In the next section, for the case $m = 1/2$, I show that if this condition of finite probability density is imposed, an explicit and simple solution is possible only for certain values of the parameters. The eigenvalues are also found in this case.\\

Based on this solution (\ref{solutions-half-S}), in the third section the problem is also solved for arbitrary non-zero $m$. An explicit expression for the eigenvalues and the allowed values of the parameters is obtained. The eigenvalue expression is very similar to the expression obtained in the case of $m = 1/2$ but the states are very different in nature. Indeed, the solutions, equations (\ref{solutions-m-S}) are so simple in functional form that it is easier to verify them by direct substitution into the angle equations (\ref{theta-full}) rather than solve the equation. Along with our earlier result  \cite{ranganathan:1} for the special case $m = 0$, this completes the solutions of the angular Chandrasekhar Page equation.  In the last section  we discuss these results.\\

\section{Solution of the equations \ref{theta-full} for $m = 1/2$} \label{Solution-half:level1} \protect

For $m = 1/2$ the angle equations  (\ref{theta-full}) become

\begin{subequations}
\label{theta-half}
\begin{eqnarray}
\frac{d}{d \theta} S_{+\frac{1}{2}} + \left[ (\frac{1}{2}) \cot\theta  + ( a \omega  \sin \theta + (\frac{1}{2}) \csc\theta ) \right] S_{+ \frac{1}{2}} & = &  ( - \lambda + a m_e \cos\theta) S_{- \frac{1}{2}}, \\
\frac{d}{d \theta} S_{-\frac{1}{2}} + \left[ (  \frac{1}{2}) \cot\theta -( a \omega  \sin \theta + (\frac{1}{2})  \csc\theta ) \right] S_{- \frac{1}{2}} & = & ( \lambda + a m_e \cos\theta) S_{+ \frac{1}{2}}. 
\end{eqnarray}
\end{subequations}
 
 We isolate the part where this equation can possibly be singular by rewriting the equations as
 
 \begin{subequations}
\label{theta-half angle}
\begin{eqnarray}
 \frac{d}{d \theta} S_{+\frac{1}{2}} + \left[ \frac{ \cot(\theta /2)}{2}  + ( a \omega \sin \theta  )  \right] S_{+ \frac{1}{2}} & = &  ( - \lambda + a m_e \cos\theta) S_{- \frac{1}{2}}, \\
\frac{d}{d \theta} S_{-\frac{1}{2}} + \left[  - \frac{ \tan (\theta /2)}{2} -( a \omega  \sin \theta ) \right] S_{- \frac{1}{2}} & = & ( \lambda + a m_e \cos\theta) S_{+ \frac{1}{2}}. 
\end{eqnarray}
\end{subequations}

The substitutions
\begin{equation}
\label{substitutions}
%\begin{eqnarray}
 S_{+\frac{1}{2}}  =  \frac{T_{+\frac{1}{2}}}{  \sin (\theta /2)} ,  \; \; \; \; \; S_{-\frac{1}{2}}  =  \frac{T_{-\frac{1}{2}}}{  \cos (\theta /2) } ,  \; \; \;  \; \; \cos \theta  =  x,  
% \end{eqnarray}
\end{equation}

 convert our equations (\ref{theta-half}) to

\begin{subequations}
\label{x eqns}
\begin{eqnarray}
\frac{d}{d x} T_{+\frac{1}{2}} - a \omega   T_{+\frac{1}{2}} & = &  \frac{  \lambda - a m_e x}{1 + x} T_{-\frac{1}{2}}, \\
\frac{d}{d x} T_{-\frac{1}{2}} + a \omega  T_{-\frac{1}{2}} & = & - \frac{\lambda + a m_e x }{1 -x} T_{+\frac{1}{2}}. 
\end{eqnarray}
\end{subequations}
These can be easily rewritten as a second order differential equation,
$$ \frac{d^2 T_{+\frac{1}{2}}}{d x^2} + \left( \frac{1}{1+x} +\frac{a m_e}{\lambda - a m_e x} \right) \frac{d T_{+\frac{1}{2}}}{d x} - \left( \frac{ \omega a }{1+x} + \frac{\omega m_e a^2 }{\lambda - a m_e x} + ( \omega a )^2 + \frac{ {\lambda}^2 - ( a m_e x )^2}{1 - x^2} \right) T_{+\frac{1}{2}} = 0. $$

The above equation gives a Sturm Liouville problem with regular singular points at $ x = \pm 1 $ and $ \lambda / (a m_e) $ and hence the solutions corresponding to acceptable wave packets can always be expressed in terms of hypergeometric functions. We confine ourselves to the case where the eigenvalue $ \left|\lambda \right| \: > \: \left| a m_e \right| $. The justification for this restriction is given in the last section. The power series solution will then converge over the entire allowed range of $x$, namely $\left[ -1 , 1 \right] $. We therefore try a series solution of the form 

\begin{equation}
\label{half power series}
T_{+\frac{1}{2}} = \sum_{n =0} A_n x^n, \; \; \; \; \; T_{-\frac{1}{2}} = \sum_{n = 0 } B_n x^n .
\end{equation}

Substituting the above in equation (\ref{x eqns}) the resulting recurrence relations for the series coefficients are,
\begin{subequations}
\label{series}
\begin{eqnarray}
n A_{n} + (n - 1 - a \omega ) A_{n-1 } - a \omega  A_{n-2} & = & \lambda B_{n-1} - a m_e B_{n-2} , \\
n B_{n} - (n - 1 - a \omega ) B_{n-1 } -  a \omega  B_{n-2} & = &  - \lambda A_{n-1} - a m_e A_{n-2} .
\end{eqnarray}
\end{subequations}

Inspection shows that
\begin{equation}
\label{B_n}
B_{n} = P (-1)^n A_{n},\;\;\; P= \pm 1.
\end{equation}

$A_n$ then satisfies

\begin{equation}
\label{A_n}
n A_n + \left(n - 1 - a \omega  + \lambda P (-1)^n \right) A_{n-1} - \left( a \omega  -a m_e (-1)^n P \right) A_{n-2} =0.
\end{equation}

This series does not converge at both the limits $\pm 1$ and thus cannot represent a wave packet over all of space. Further convergence of the series for $T_{+ \frac{1}{2}}$ alone is not enough as the function of actual interest $S_{+ \frac{1}{2}}$ has a $\sin \left( \theta / 2 \right) $ appearing in the denominator, equation (\ref{substitutions}). To overcome the divergence due to this sine, it is necessary to have $A_{2 n} = - A_{2 n + 1}$. Imposition of this condition leads to determination of the eigenvalues $\lambda$ and solutions, which are possible only for
\begin{subequations}
\label{solution-half}
\begin{eqnarray}
\lambda & = & -P + a m_e, \\
a \, \omega & = & - P a m_e, \\
A_1 & = & - A_0, \\
A_n & = & 0, \; \; n > 1.
\end{eqnarray}
\end{subequations}

The solutions for $ m = 1/2 $ are
\begin{subequations}
\label{solutions-half-S}
\begin{eqnarray}
S_{+\frac{1}{2}} & = & - \frac{\sin \left( \theta / 2 \right)}{\sqrt{2}} \\
S_{-\frac{1}{2}} & = & P \frac{\cos \left( \theta / 2  \right)}{\sqrt{2}} .
\end{eqnarray}
\end{subequations}

These wave packets have their maxima at one pole of the black hole and their minima at the other.\\

\section{Solution of the equations \ref{theta-full} for arbitrary $m = k+1/2$} \label{Solution-arb:level1} \protect

For arbitrary $m = k + 1/2$ the angle equations  (\ref{theta-full}) become

\begin{subequations}
\label{theta-m}
\begin{eqnarray}
\frac{d}{d \theta} S_{+\frac{1}{2}} + \left[ (\frac{1}{2}) \cot \theta  + (a \omega  \sin \theta +  (k + \frac{1}{2}) \csc \theta ) \right] S_{+ \frac{1}{2}} & = &  ( - \lambda + a m_e \cos\theta) S_{- \frac{1}{2}}, \\
\frac{d}{d \theta} S_{-\frac{1}{2}} + \left[ (  \frac{1}{2}) \cot \theta -( a \omega  \sin \theta +  (k + \frac{1}{2}) \csc \theta ) \right] S_{- \frac{1}{2}} & = & ( \lambda + a m_e \cos\theta) S_{+ \frac{1}{2}}. 
\end{eqnarray}
\end{subequations}
 
 As in the previous case, we use the substitutions given in equation (\ref{theta-half angle}) and equation (\ref{substitutions}) to obtain

\begin{subequations}
\label{xm eqns}
\begin{eqnarray}
\frac{d}{d x} T_{+\frac{1}{2}} - \left( a \omega   + \frac{k}{(1 - x^2)} \right)T_{+\frac{1}{2}} & = &  \frac{  \lambda - a m_e x}{1 + x} T_{+\frac{1}{2}}, \\
\frac{d}{d x} T_{-\frac{1}{2}} + \left( a \omega   + \frac{k}{(1 - x^2)} \right) T_{-\frac{1}{2}} & = & - \frac{\lambda + a m_e x }{1 -x} T_{+\frac{1}{2}}. 
\end{eqnarray}
\end{subequations}

These can be easily rewritten as a single second order differential equation with regular singular points at $ x = \pm 1 $ and $\lambda / (a m_e) $. Once again we confine ourselves to the case where the eigenvalue $ \left|\lambda \right| \: > \: \left| a m_e \right| $. The power series solution will then converge within the entire allowed range of $x$, namely $\left[ -1 , 1 \right] $. As before, we try a solution of the form,
\begin{equation}
\label{m power series}
T_{+\frac{1}{2}} = \sum_{n =0} C_n x^n, \; \; \; \; \; T_{-\frac{1}{2}} = \sum_{n = 0 } D_n x^n .
\end{equation}

On substituting the above in equation (\ref{xm eqns}), the resulting recurrence relations for the series coefficients are,
\begin{subequations}
\label{series-m}
\begin{eqnarray}
n C_{n} - (k + a \omega ) C_{n-1 } -  (n - 2) C_{n-2} + a \omega  C_{n-3}& = & \lambda D_{n-1} - (a m_e + \lambda ) D_{n-2} + a m_e D_{n-3} , \\
n D_{n} + (k + a \omega ) D_{n-1 } -  (n - 2) D_{n-2} - a \omega  D_{n-3}& = &  - \lambda C_{n-1} - (a m_e + \lambda ) C_{n-2} - a m_e C_{n-3}.
\end{eqnarray}
\end{subequations}

Again the second equation of the pair becomes an identity if
\begin{equation}
\label{D_n}
D_{n} = P (-1)^n C_{n},\;\;\; P= \pm 1.
\end{equation}

$C_n$ then satisfies

\begin{equation}
\label{C_n}
n C_n -\left(k + a \omega  - P \lambda (-1)^n \right) C_{n-1} + \left(n - 2 - (a m_e + \lambda P (-1)^n) \right) C_{n-2} - \left( a \omega  + a m_e (-1)^n P \right) C_{n-3} =0.
\end{equation}

This series does not converge at both the limits $\pm 1$ and thus cannot represent a wave packet over all of space. Further convergence of this series for $T_{\frac{1}{2}}$ alone is not enough as the function of actual interest $S_{+ \frac{1}{2}}$ has a $\sin \left( \theta / 2 \right) $ appearing in the denominator, equation (\ref{substitutions}). To overcome the divergence due to the sine, it is necessary to have $C_{2 n} = - C_{2 n + 1}$. Imposition of this condition leads to determination of the eigenvalues $\lambda$ and solutions, which are possible only for
\begin{subequations}
\label{solution-m}
\begin{eqnarray}
\lambda & = & -P + a m_e -P k, \\
a \, \omega & = & - P a m_e, \\
k & = & 2p, \; \; \; p = 0, 1,\: 2,\: 3 \: \ldots \\
C_{2n} & = & \frac{ (-1)^n p !}{n ! (p-n)! }, \; n < p,\\
C_{2n} & = & 0, \; n \geq p.
\end{eqnarray}
\end{subequations}

It is interesting to note that unlike the flat space case, only positive values of $k$ are now allowed. That is only corotating wave packets are possible. The solutions for $ m = k + 1/2 $ are
\begin{subequations}
\label{solutions-m-S}
\begin{eqnarray}
S_{+\frac{1}{2}} = - \frac{1}{2^{m}} \sqrt{\frac{(2 k + 1 )!}{(k + 1 )! \; k!}} \sin \left( \theta / 2 \right)  (\sin \theta )^k ,\\
S_{-\frac{1}{2}} =    \frac{P}{2^{m}} \sqrt{\frac{(2 k + 2 )!}{(k + 1 )! \; k!}} \cos \left( \theta / 2 \right) (\sin \theta )^k.
\end{eqnarray}
\end{subequations}\\

These go over to the asymptotic forms found in \cite{batic:2} at the poles and have their maxima close to the equatorial plane.\\

\section{Discussion} \label{discussion:level1} \protect

The solutions given in equations (\ref{solutions-half-S}) and (\ref{solutions-m-S}) are so simple in form that the it seems surprising that they have not been obtained before. The reason appears to be that they are not easily expressed in series composed of the spin weighted spheroidal harmonics requiring an infinite number of these terms for their representation. Thus previous efforts were using an ill adapted basis and consequently did not obtain an easy convergence. In our case the factoring out of the half angle term is responsible for the quick convergence of the remaining part.\\

There is a simple physical explanation for the occurrence of the half angles. The action of the spin weight $1/2$ raising and lowering operators \cite{newman:1, goldberg:1} on these states is as follows.
\begin{subequations}
\label{spin-ladder}
\begin{eqnarray}
\left[ \frac{d}{d \theta} + \frac{1}{2} \left( \cot \theta + \csc \theta \right) \right] \sin ( \theta / 2 ) & = & \cos (\theta /2) \; = \;  _{-\frac{1}{2}}Y_{0 \: 0}, \\
\left[ \frac{d}{d \theta} + \frac{1}{2} \left( \cot \theta - \csc \theta \right) \right] \cos ( \theta / 2 ) & = &  - \sin (\theta /2) \; = \; _{\frac{1}{2}}Y_{0 \: 0}, \\
\left[ \frac{d}{d \theta} + \frac{1}{2} \left( \cot \theta - \csc \theta \right) \right] \sin ( \theta / 2 ) &  = & -  \frac{\cos \theta }{2 \cos ( \theta /2)},\\
\left[ \frac{d}{d \theta} + \frac{1}{2} \left( \cot \theta + \csc \theta \right) \right] \cos ( \theta / 2 ) &  = & \frac{\cos \theta }{2 \sin ( \theta /2)}. 
\end{eqnarray}
\end{subequations}

Such states with $ s > m$ are explicitly excluded in the definition of the spin weighted harmonics, as they are defined as functions of spin and {\it{total angular momentum}} \cite{newman:1, goldberg:1}. In the language of flat space quantum mechanics, the solution expanded in terms of the spheroidal states  are given in the basis $\left|J,S\right\rangle$. The states as expanded here are in the basis $\left|L,S\right\rangle$. That is, we conceive of states like $ _{\frac{1}{2}}Y_{0 \: 0}$ as a product of a spin and a {\it{orbital angular momentum}} state. Both sets of course span the same space and will yield the same results, only the second set happens to be more efficient in this situation.\\

Consider first the eigenvalues we have found. Our neglect of all eigenvalues $ \left| \lambda \right| < \left| a m_e \right| $ is for the following reason. In the classical domain, these correspond to trajectories with impact parameter less than $a$, thus they will penetrate deep inside the horizons and are expected to be captured. In quantum terms, the spin contribution of $\hbar /2$ alone is many orders of magnitude greater than the even extremal values of $a$ for any particle with mass less than a picogram and thus the neglect of states with $\left| \lambda \right| < \left|a m_e \right|$ is justified for all known Dirac particles.\\

It is interesting to note that the eigenvalues given in equations (\ref{solution-half}a) and (\ref{solution-m}a) are the linear sum of the eigenvalue found by Dirac for the flat space central force problem along with a contribution which is the product of the mass $m_e$ of the particle and $a$ the specific angular momentum of the hole. Due to frame dragging we expect that the particle will indeed acquire such an additional contribution to the angular momentum as seen by a distant observer.\\

 There is one important difference in the eigenvalue spectrum from the flat space case. In that situation $k = \pm (j + 1/2)$ and for a given $j$ all permitted values of the azimuthal quantum number $-j \leq m \leq j$ can occur. Here the only value which occurs is $m = j$. That is the hole ensures that all allowed states with non zero orbital angular momentum are in the equatorial plane and are co-rotating. Further the total angular momentum of the Dirac particle is aligned with that of the hole $m = j$. This is to be expected on classical grounds.\\

 The separation of successive eigenvalues by two units of angular momentum is also to be expected given the spin 2 nature of the gravitational field. It is perturbations of this field which will cause transitions between the states of different angular momenta. This is analogous to the situation in the case of atoms where the angular momentum states are separated by one unit of angular momentum, corresponding to the spin 1 nature of the electromagnetic field. This causes the spectrum of states for an atom to contain only states of half integral angular momenta or only integral angular momenta.\\

 Next consider the states.  The maxima for all the states with $m \neq 1/2$ are very close to the equatorial plane and converge very rapidly to it with increasing $m$. Equally interesting is the fact that this displacement from the equatorial plane is different for spin up and spin down states. The spin up states have their maxima below the equator while the spin down states have their maxima above the equatorial plane. This is to be expected on the basis of the magnetic and gravimagnetic interactions of the Dirac particle in the non uniform field of the hole.\\
 
 The states with $m = 1/2$ are exceptional as these alone have their maxima at the poles. Again with spin up at the south pole and spin down at the north pole. Along with the confinement of the other states with non-zero orbital angular momentum, this is strongly reminiscent of the accretion disk and polar emission of classical particles orbiting a black hole. Of course, until the radial Chandrasekhar Page equations are also solved this is only a analogy.\\

The radial Chandarsekhar Page equations can be brought to almost the same form as the angle equations studied here and thus may be amenable to solution by the same method.

\end{document}